\documentclass[twocolumn,english,aps,prd,superscriptaddress,longbibliography]{revtex4-1}
\usepackage[T1]{fontenc}
\usepackage{color}
\usepackage{amsmath}
\usepackage{amssymb}
\usepackage{graphicx}
\usepackage{natbib}
\usepackage{braket}
\usepackage{psfrag}
\usepackage{tikz}
\usepackage{physics}
\usepackage{cancel}
\usepackage{bbm}
\usepackage{float}
\usepackage[colorlinks=true,urlcolor=blue,citecolor=blue,linkcolor=blue]{hyperref}
\usepackage[normalem]{ulem}

\usepackage{xcolor}
\definecolor{mjhcolor}{rgb}{0.27, 0.01, 0.56}
\definecolor{swcolor}{rgb}{0.18, 0.70, 0.18}
\definecolor{fedcolor}{rgb}{0.6, 0.2, 0.2}

\usepackage{array,booktabs}
\newcolumntype{M}[1]{>{\centering\arraybackslash}m{#1}}

\begin{document}

\title{Direct implementation of a perceptron in superconducting circuit quantum hardware}

\author{Marek Pechal}
\affiliation{IBM Quantum, IBM Research Europe - Zurich, S\"{a}umerstrasse 4, 8803 R\"{u}schlikon, Switzerland}
\affiliation{ETH Zurich, Department of Physics, 8093 Z\"urich, Switzerland}

\author{Federico Roy}
\affiliation{IBM Quantum, IBM Research Europe - Zurich, S\"{a}umerstrasse 4, 8803 R\"{u}schlikon, Switzerland}
\affiliation{Theoretical Physics, Saarland University, 66123 Saarbr\"ucken, Germany}
\affiliation{Walther-Mei{\ss}ner-Institut, Bayerische Akademie der Wissenschaften, 85748 Garching, Germany}

\author{Samuel A. Wilkinson}
\affiliation{Department of Physics, Friedrich-Alexander-Universit\"at Erlangen-N\"urnberg (FAU), Staudtstr. 7, 91058 Erlangen, Germany}

\author{Gian Salis}
\affiliation{IBM Quantum, IBM Research Europe - Zurich, S\"{a}umerstrasse 4, 8803 R\"{u}schlikon, Switzerland}

\author{Max Werninghaus}
\affiliation{IBM Quantum, IBM Research Europe - Zurich, S\"{a}umerstrasse 4, 8803 R\"{u}schlikon, Switzerland}
\affiliation{Walther-Mei{\ss}ner-Institut, Bayerische Akademie der Wissenschaften, 85748 Garching, Germany}
\affiliation{Physik-Department, Technische Universit\"{a}t M\"{u}nchen, 85748 Garching, Germany}

\author{Michael J. Hartmann}
\affiliation{Department of Physics, Friedrich-Alexander-Universit\"at Erlangen-N\"urnberg (FAU), Staudtstr. 7, 91058 Erlangen, Germany}
\affiliation{Max Planck Institute for the Science of Light, 91058 Erlangen, Germany}

\author{Stefan Filipp}
\affiliation{IBM Quantum, IBM Research Europe - Zurich, S\"{a}umerstrasse 4, 8803 R\"{u}schlikon, Switzerland}
\affiliation{Walther-Mei{\ss}ner-Institut, Bayerische Akademie der Wissenschaften, 85748 Garching, Germany}
\affiliation{Physik-Department, Technische Universit\"{a}t M\"{u}nchen, 85748 Garching, Germany}
\affiliation{Munich Center for Quantum Science and Technology (MCQST), Schellingstra\ss e 4, 80799 M\"{u}nchen, Germany}

\date{\today}

\begin{abstract}
The utility of classical neural networks as universal approximators suggests that their quantum analogues could play an important role in quantum generalizations of machine-learning methods. Inspired by the proposal in \cite{Torrontegui2019}, we demonstrate a superconducting qubit implementation of an adiabatic controlled gate, which generalizes the action of a classical perceptron as the basic building block of a quantum neural network. 
We show full control over the steepness of the perceptron activation function, the input weight and the bias by tuning the adiabatic gate length, the coupling between the qubits and the frequency of the applied drive, respectively.
In its general form, the gate realizes a multi-qubit entangling operation in a single step, whose decomposition into single- and two-qubit gates would require a number of gates that is exponential in the number of qubits. Its demonstrated direct implementation as perceptron in quantum hardware may therefore lead to more powerful quantum neural networks when combined with suitable additional standard gates.
\end{abstract}

\maketitle

\section{Introduction}

Artificial neural networks and engineered quantum systems are both quickly developing technologies with far-reaching potential applications. 
The promise that quantum computing can solve certain problems exponentially faster than classical computing technology and the ever growing thirst for computational power of machine learning applications has triggered substantial interest in the development of quantum machine learning methodology \cite{Biamonte2017,Rocchetto2018,Sarma2019,Zoufal2019,Melko2019,Abbas2021,Havlivcek2019,Glick2021}. Whereas some work explored the acceleration of specific computational tasks in machine learning with quantum encodings \cite{HHL}, a major part of the research considers quantum neural networks as counterparts to artifical neural networks in classical software. Quantum neural networks are largely implemented as variational quantum circuits \cite{Farhi2014,Peruzzo2014, Moll2018, Kandala2017}, that are composed of parametrized gates and where finding optimal parameters corresponds to the training of the network. Moreover, it has been shown that quantum speed-up is possible in supervised machine learning \cite{Liu2021}, providing a perspective for hardware-efficient machine learning realizations.

An important question for quantum neural networks is their expressivity, i.e. which mappings between input and output states can be realized by the network. In the classical domain, expressivity is guaranteed by the universal approximation theorem, which requires a nonlinear activation function for the perceptrons in the network. In the quantum setting, it has been shown that similar notions of universal approximation exist for functions mapping gate parameters to the state prepared by the quantum circuit \cite{Perez2021}.
However, currently there is no universally accepted way to extend the concept of classical neural networks to quantum systems \cite{Behrman2000,Schuld2014}. One option is to design a quantum perceptron as a unitary whose action on a set of basis states matches a classical perceptron \cite{Torrontegui2019}. Such a building block can serve as a tool to enable experimental studies and development of quantum neural networks. Since such an approach includes the functionality of classical neural networks in limiting cases, it has the expressive power guaranteed by the universal approximation theorem in those cases and thus forms a natural starting point for exploring quantum generalizations. Another advantage of this approach is that the perceptron may be directly realized at the hardware level, rather than encoded in software: this neuromorphic approach is more hardware efficient and may thus offer significant advantages for scaling the concept to larger and more data intensive applications.

\begin{figure}
\includegraphics[width=6.5cm]{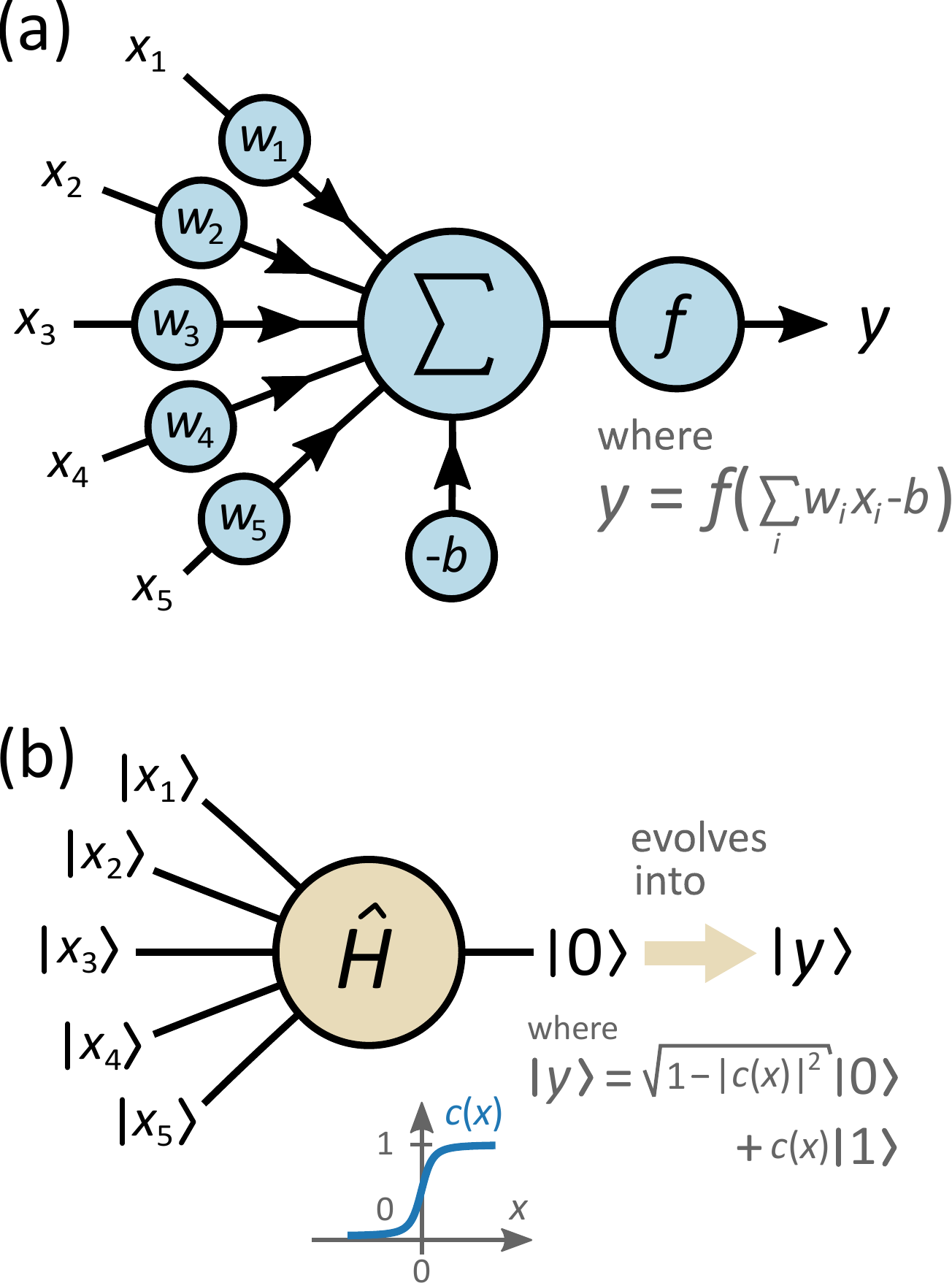}
\caption{(a) Schematic representation of a classical perceptron with $n=5$ inputs. The perceptron output $y$ is given by a non-linear activation function $f(\cdot)$ applied to the sum of input values $x_1,...,x_n$ individually weighted by $w_1,...,w_n$ and biased by a term $b$. A quantum analog (b) can be constructed, where the inputs are encoded in quantum states $\ket{x_1}, ..., \ket{x_n}$ of $n$ qubits and an output qubit undergoes a rotation by an angle of $\theta = \arcsin[c(x)]$ determined by the inputs values, leading to an excited population of $|c(x)|^2$ when starting the output qubit in the ground state.
}\label{fig:networks}
\end{figure}

The construction of the quantum counterpart to a perceptron is easily illustrated with the example of a binary classifier perceptron.  This simple classical perceptron takes a number of binary inputs $x_i \in \{0,1\}$, each with an associated weight $w_i$, and outputs another binary value $y$ depending on whether the linear combination $x_\text{in} = w_1 x_1 + \ldots$ exceeds a given threshold or not:
\begin{align*}
y = \left\{\begin{array}{l}
    0\text{ if } x_\text{in} + b < 0\\
    1\text{ if } x_\text{in} + b \ge 0\\
\end{array}\right.,
\end{align*}
or, written compactly in terms of the Heaviside step function, $y = \Theta(x_\text{in}+b)$ with an additional bias $b$. More generally, as illustrated in Fig.~\ref{fig:networks}(a), the step function may be replaced by a continuous activation function $f$ and the inputs and outputs promoted from binary values to continuous ones. The key advantage of such continuous activation functions is that they have finite gradients which can be used for training the network in gradient descent approaches.

In a quantum generalization of the classifier perceptron, the binary variables can be naturally represented by qubit states, which we will label as $|0\rangle$ and $|1\rangle$. The action of the perceptron can then be represented by a gate $U$ whose effect on the output qubit depends on the states of the input qubits:
\[
  |x_1 x_2\ldots\rangle_{\mathrm{in}} \otimes |0\rangle_{\mathrm{out}} \xrightarrow{U}
  \left\{\begin{array}{l}
  |x_1 x_2\ldots\rangle_{\mathrm{in}} \otimes |0\rangle_{\mathrm{out}}\\
  \text{ if } x_\text{in} + b \equiv \sum_{j} w_j x_j + b < 0\\
  \\
  |x_1 x_2\ldots\rangle_{\mathrm{in}} \otimes |1\rangle_{\mathrm{out}}\\
  \text{ if } x_\text{in} + b \equiv \sum_{j} w_j x_j + b \ge 0\\
  \end{array}\right.
\]
When the input is a product state $|x_1 x_2\ldots\rangle_{\mathrm{in}}$, then the unitary is simply a rotation of the output qubit by $0$ or $\pi$ depending on the sign of $x_\text{in} + b$ \cite{Cao2017}. This concept of a quantum perceptron can be further generalized, similarly to the classical case, by letting the rotation angle have a different dependence on $x$ than a simple Heaviside function. The action of the perceptron, shown schematically in Fig.~\ref{fig:networks}(b), is then described as
\begin{align}
  |x_1x_2 \ldots\rangle_{\mathrm{in}} \otimes |0\rangle_{\mathrm{out}} &\xrightarrow{U}
  |x_1x_2 \ldots\rangle_{\mathrm{in}} \otimes |
  y
  \rangle_{\mathrm{out}},\nonumber\\
  \text{where }
  |y\rangle_{\mathrm{out}} &= \sqrt{1-|c(x)|^2}|0\rangle + c(x)|1\rangle.\label{eq:perceptrongate}
\end{align}
Here the excitation amplitude $c(x)$ is a continuous, step-like function of $x\equiv x_\text{in}+b$ such as depicted in Fig.~\ref{fig:networks}. The previously described binary perceptron is the special case where $c(x)$ is the Heaviside step function $\Theta(x)$.

Since the perceptron gate is a unitary operation, it can in principle be implemented via any universal set of quantum gates, such as single-qubit rotations and controlled-NOT (CNOT) gates. However, the depth of such a decomposition in general grows exponentially in the number of input qubits. Therefore, we instead realize the perceptron gate directly, making use of an adiabatic protocol. This approach allows us to implement the gate with a single adiabatic pulse whose duration does not scale with the number of input qubits: From Eq.~(\ref{eq:perceptrongate}) it can be seen that the perceptron acts like a multi-qubit controlled gate. However, instead of triggering an operation on the target only when the control register is in the $\ket{1\dots 1}$ state (like a Toffoli gate \cite{Fredkin1982}, which applies an $X$ gate to the target only when the control qubits are in the state $\ket{11}$), the perceptron gate applies a different operation to the output qubit for each possible input basis state $\ket{x_1x_2\dots}_\textrm{in}$.
As the number of input basis states grows exponentially in the number of qubits, a standard decomposition into elementary gates leads to an expontial growth of the circuit depth (see Section~\ref{sec:equiv-circuit}), which is in contrast to the adiabatic implementation discussed in the following.

\section{The perceptron gate}\label{sec:perceptrongate}
\subsection{Concept}\label{subsec:concept}

We construct the perceptron unitary with an approach that is a slight modification of a theoretical proposal by Torrontegui and Garc\'{i}a-Ripoll \cite{Torrontegui2019}, as described below. An implementation of a similar operation was also proposed \cite{Cao2017} based on repeated measurements and feedback, while a gate-based approach to construction of activation functions was demonstrated in \cite{Yan2020}.

\begin{figure}[t]
    \centering
    \includegraphics[width=8.0cm]{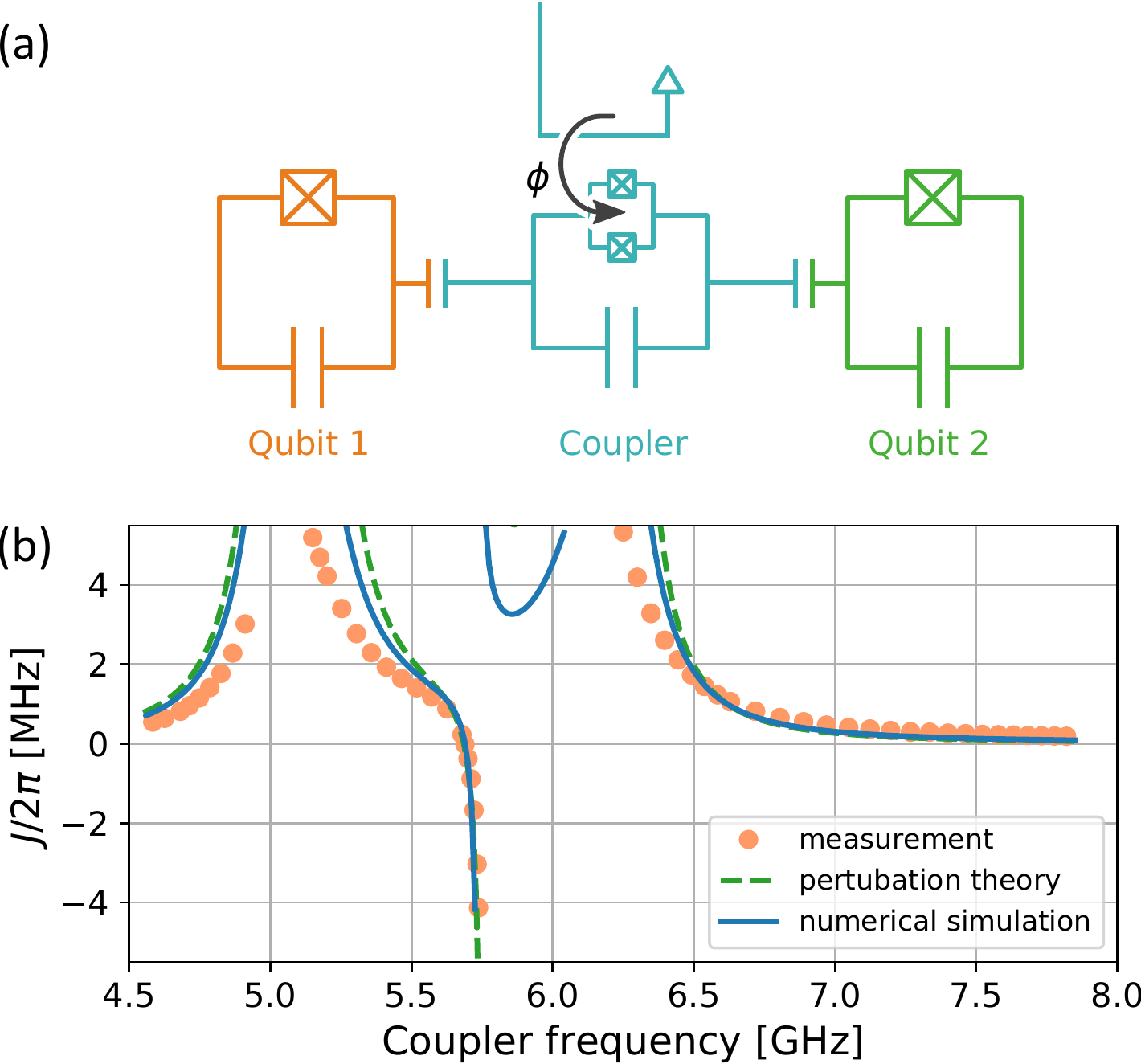}
    \caption{(a) Diagram of the two-qubit experimental setup: two fixed frequency transmons are capacitively coupled via a flux-tunable coupler whose frequency determines the ZZ coupling between the qubits. (b) Measured value of the ZZ coupling as a function of the tunable coupler's frequency (orange dots). Here we plot the coupling strength $J$ conventionally defined via $H_{\mathrm{int}} = J |11\rangle\langle 11|$. In the perceptron context, the weight $w$ from Eq.~(\ref{eq:zzintham}) is given by $w = -J$. The solid line is the result of numerically diagonalizing the hamiltonian of the system and the dashed line is the result of the lowest order perturbative calculation (Appendix \ref{app:pertThJ}).}
    \label{fig:coupler}
\end{figure}

Our protocol, which we implement in a device with two fixed-frequency superconducting transmon qubits \cite{Koch2007} interacting via a tunable coupler \cite{McKay2016} (schematically depicted in Fig.~\ref{fig:coupler}(a)) is based on an adiabatic unitary evolution applied to the output qubit. This gate $U_{\mathrm{ad}}$ is designed in such a way that the final excited state amplitude depends on the detuning $\Delta$ of a microwave drive frequency from the output qubit frequency via the step-like response function $c(\Delta)$:
\begin{equation}\label{eq:1qgate}
  |0\rangle_{\mathrm{out}}
  \xrightarrow{U_{\mathrm{ad}}}
  \sqrt{1-|c(\Delta)|^2}|0\rangle_{\mathrm{out}} + c(\Delta)|1\rangle_{\mathrm{out}}
\end{equation}
Note that this becomes exactly the mapping in Eq.~(\ref{eq:perceptrongate}) if the detuning $\Delta$ is equal to the linear combination of inputs $\sum_j w_j x_j + b$. 
We achieve this linear dependence of $\Delta$ on the input states by subjecting the system to a $ZZ$ interaction between each of the input qubits and the perceptron qubit:
\begin{equation}\label{eq:zzintham}
  H_{\mathrm{int}} = -\sum_j w_j
  |1\rangle\langle 1|_j \otimes |1\rangle\langle 1|_{\mathrm{out}}
\end{equation}
Here $j$ enumerates the input qubits and "$\mathrm{out}$" labels the output qubit. If the input qubits are in a product state $|x_1\ldots\rangle$, this is equivalent to shifting the frequency of the output by $-\sum_j w_j x_j$, or $-w_1 x_1$ in the case of a single input.

\begin{figure}[t]
\includegraphics[width=8.0cm]{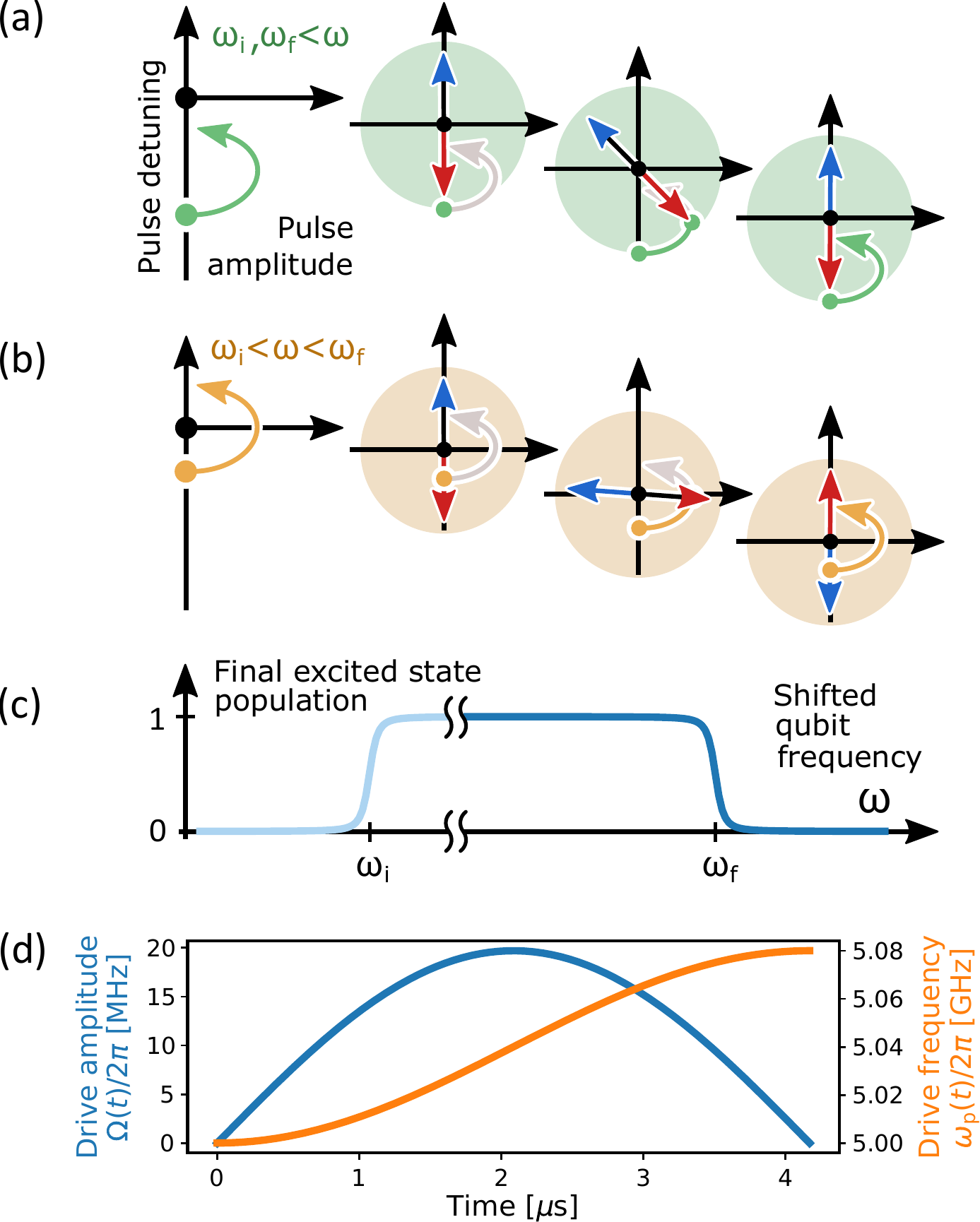}
\caption{
(a,b) Illustration of the effect of an adiabatic chirped pulse with initial and final frequencies $\omega_{\mathrm{i}}$ and $\omega_{\mathrm{f}}$ on a qubit with a frequency $\omega$. (a) If the initial and final detunings have the same sign, the basis states $|0\rangle$ and $|1\rangle$ are unchanged (up to a phase). (b) If the signs are opposite, the states get swapped. This results in a dependence of the final state on the shifted qubit frequency shown in (c). The perceptron activation function is realized by this dependence in the vicinity of $\omega\approx\omega_{\mathrm{f}}$ (darker line). (d) Time-dependence of the amplitude $\Omega(t)$ and frequency $\omega_p(t)$ of an exemplary adiabatic chirped pulse realizing the output-qubit rotation.}\label{fig:protocol}
\end{figure}

\begin{figure}
\includegraphics[width=7.7cm]{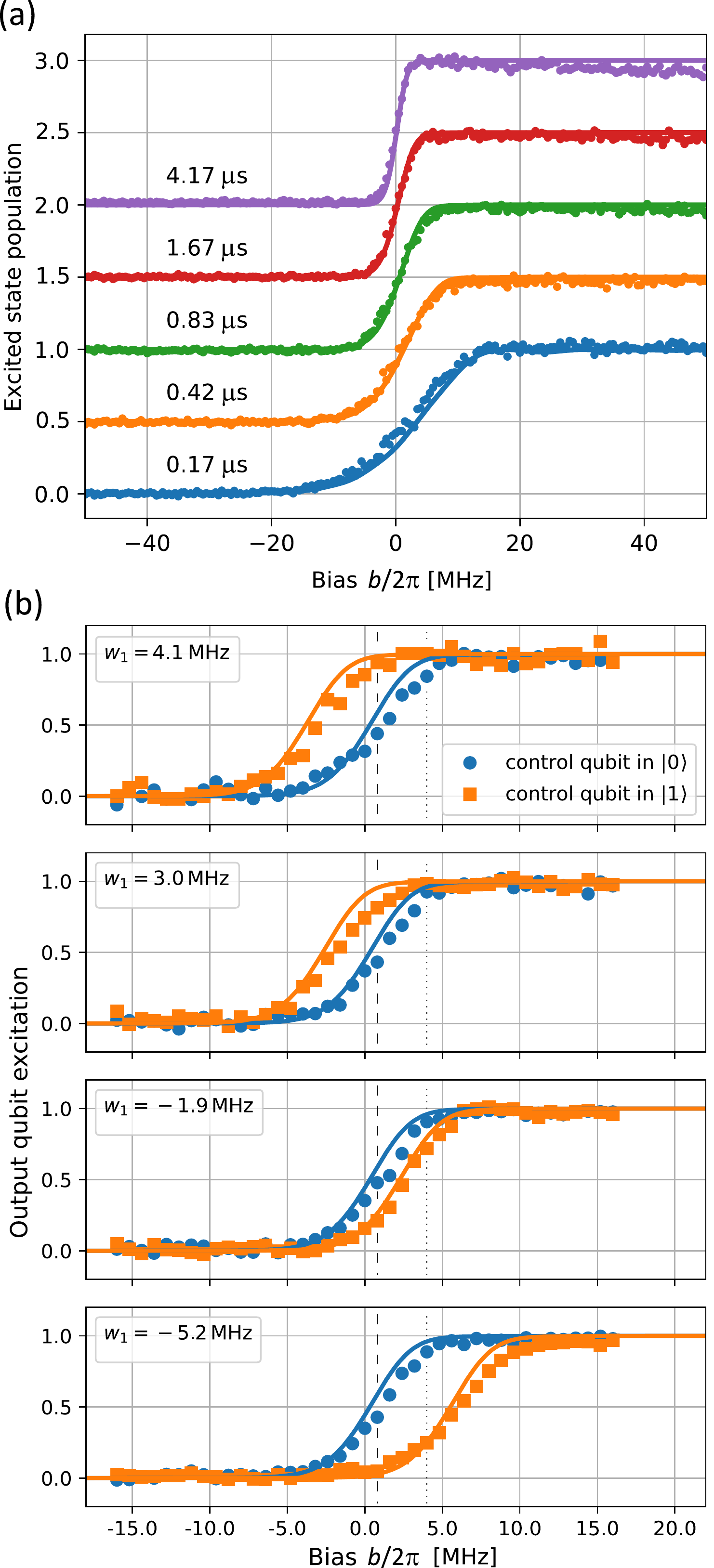}
\caption{(a) The quantum perceptron activation functions: The excited state population $p_{\mathrm{e}}$ of the output qubit as a function of its final detuning $\Delta$ after an adiabatic chirped pulse is applied on the ground state of two-qubit system. Results for different pulse lengths $T$ are offset for visualization. The characteristic width of the activation function is inversely proportional to the pulse lengths $T$. The solid lines result from a simulation of unitary evolution under the adiabatic drive, with no fit parameters in this model. (b) Frequency shift of the activation function dependent on the state of the input qubit, for a pulse length of $T=1.67\,\mu\mathrm{s}$ Both the magnitude and sign of this shift can be varied by tuning the frequency of the coupler. In these plots, the output qubit excitations are rescaled to correct for relaxation and readout imperfections, in order to highlight the agreement of the curve's shape with the simulation results. The frequency spacing between the simulated curves is not obtained by fitting but rather predicted from measured $J$ values (see Fig.~\ref{fig:coupler}(b)). The vertical lines indicate values of $b/2\pi = 0.8\,\mathrm{MHz}$ (dashed) and $b/2\pi = 4.0\,\mathrm{MHz}$ (dotted) which are used in Fig.~\ref{fig:maxdiff}(a).}
\label{fig:scurves}
\end{figure}

\subsection{Implementation}\label{subsec:implementation}

The ZZ interactions need to be configurable in-situ to allow training of the network. This can be achieved by using tunable couplers \cite{McKay2016} mediating interactions between the output qubit and each input qubit. When the couplers are sufficiently far detuned from the qubits, the dispersive approximation is valid and the interactions effectively introduce $ZZ$ coupling terms, giving rise to the Hamiltonian in Eq.~(\ref{eq:zzintham}). The individual interaction strengths representing the network weights can be tuned by changing the frequencies of the couplers \cite{Li2020}.
Using this scheme we can control the weight of each input by tuning of the respective coupler frequency and the bias point by changing the frequency of the microwave drive.

In the two-qubit device used for our experimental demonstration, qubit 1 at a frequency $\omega_1/2\pi = 6.189\,\mathrm{GHz}$ serves as the output while qubit 2 at a frequency $\omega_2/2\pi = 5.089\,\mathrm{GHz}$ is the single-valued input register. The qubits have anharmonicities  $\alpha_{1}/2\pi=-286\,\mathrm{MHz}$ and $\alpha_{2}/2\pi=-310\,\mathrm{MHz}$. The coupler has a frequency tunable from its maximum of approximately $\omega_{\mathrm{c},\mathrm{max}}/2\pi = 7.8\,\mathrm{GHz}$ to well below the frequencies of both qubits. The qubit-coupler interaction strengths are approximately $g_{1}/2\pi=142\,\mathrm{MHz}$ and $g_{2}/2\pi=116\,\mathrm{MHz}$.

Using Ramsey-type measurements with the input qubit either in $|0\rangle$ or in $|1\rangle$, we characterize the dependence of the ZZ coupling on the frequency of the coupler and observe that it can be tuned over a range of a few MHz (see Fig.~\ref{fig:coupler}), covering both positive and negative values. The main features of the measured dependence are well reproduced by an analytical expression obtained from 4\textsuperscript{th} order perturbation theory (see Eq.~(\ref{eq:pertThJ})), and similarly by the numerical diagonalization of the Hamiltonian (see Eq.~(\ref{eq:ham})).

The single-qubit gate, expressed by Eq.~(\ref{eq:1qgate}), which underlies the quantum perceptron protocol is realized by applying an adiabatic chirped pulse to qubit 1 \cite{Roth2019, Salis2020}. The pulse's initial frequency $\omega_{\mathrm{i}}$ is far below the output qubit's frequency $\omega_1$ (where $\omega_1$ is defined in the absence of the interaction $H_{\mathrm{int}}$) while its final frequency $\omega_{\mathrm{f}}$ is detuned from it by $b \equiv \omega_{\mathrm{f}} - \omega_1$. This means the initial detuning of the drive from the qubit is negative while the final detuning is $\Delta = \sum_j w_j x_j + b$. Note that the bias $b$, being equal to the final chirp detuning from the unshifted qubit frequency, can be tuned arbitrarily by changing $\omega_{\mathrm{f}}$.

Under perfect adiabatic conditions, as illustrated in Fig.~\ref{fig:protocol}(a,b), if the initial and the final detuning have the same sign, the two basis states $|0\rangle$ and $|1\rangle$ remain unchanged by the pulse (up to accumulated phases). If they have opposite signs, the states are flipped. The resulting dependence of the final excited state population on the qubit frequency is illustrated in Fig.~\ref{fig:protocol}(c). Since we choose the initial frequency of the chirped pulse $\omega_{\mathrm{i}}$ significantly lower than the lowest possible frequency of the output qubit ($\omega_1 - \sum_{j,w_j>0} w_j$), we neglect the rising edge of the function. The adiabatic operation is then exactly described by Eq.~(\ref{eq:perceptrongate}) with $|c(\Delta)|^2 = \Theta(\Delta)$ and $\Delta = \sum_j w_j x_j + b$. Smooth response functions $|c(\Delta)|^2$ arise naturally from imperfect adiabaticity of the chirp pulse: With a pulse of finite duration $T$, the process becomes non-adiabatic when the detuning $\Delta$ is on the order of $1/T$ or smaller. This leads to a smoothening of the step in the response function and a finite width of roughly $1/T$.
In our experiment, the chirped pulse has a time-dependent frequency $\omega_{\mathrm{p}}(t) = \omega_{\mathrm{i}} + (\omega_{\mathrm{f}}-\omega_\mathrm{i})\sin^2(\pi t/2 T)$ and amplitude $\Omega(t) = \Omega_0 \sin(\pi t/T)$, (Fig. \ref{fig:protocol}d). As a time-transformed version of a hyperbolic secant pulse \cite{Hioe1984, Baum1985}, for which analytical solutions of the population transfer exist, it leads to hyperbolic shaped activation functions with few fitting parameters (see Appendix \ref{app:chirped}).

Fig.~\ref{fig:scurves}(a) shows the measured effect of the pulse on the qubit state as a function of the bias $b = \omega_{\mathrm{f}} - \omega_1$ for several durations, $T$, of the pulse. Since the control qubit is left in its ground state, i.e. $x_1 = 0$, the curves are independent of the weight $w_1$. In these measurements, performed with a pulse amplitude $\Omega_0/2\pi = 19.7\,\mathrm{MHz}$ and initial frequency $\omega_{\mathrm{i}}=\omega_{\mathrm{f}}-80\,\mathrm{MHz}$, the qubit is initialized in its ground state and its excited state population $p_{\mathrm{e}}$ after the pulse is observed to follow a step-like curve, where the slope of the transition increases with the pulse length. The broadening of the response curve due to non-adiabatic effects is well reproduced by unitary simulations, where we simply evolve the driven two-level-system Hamiltonian in the rotating frame, taking the rotating wave approximation. We take $T=1.67\rm{\mu s}$ to be the default gate time for the following results.

The effect of the ZZ coupling between the output and the input qubit is a shift of the S-shaped activation function dependent on the input qubit being in the excited state, as demonstrated in Fig.~\ref{fig:scurves}(b). The curves corresponding to the input qubit being in the ground state (blue) are unshifted, while the ones for an excited input qubit (orange) show a shift which is adjustable by changing the coupler frequency. This shift equals the ZZ coupling strength $ J=-w_1$ (where the sign flip is included in the definition of the weight such that the activation function is increasing, as per convention) in the single connection between the output and the input qubit in the quantum perceptron gate. Depending on the relative frequency of the coupler with respect to the qubit frequencies, the weight can be made positive or negative (see Fig.~\ref{fig:coupler}).

\begin{figure}
\includegraphics[width=8.5cm]{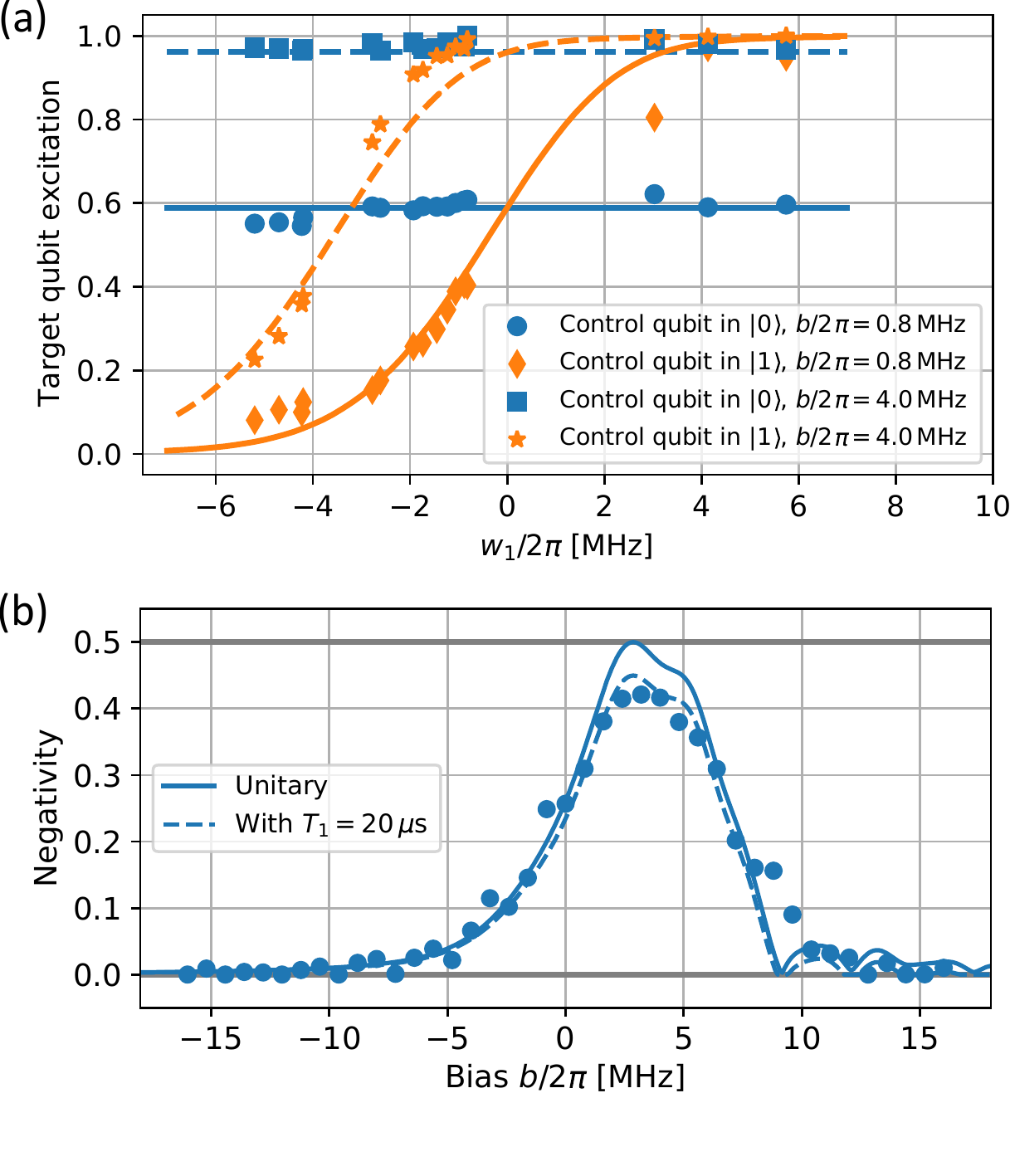}
\caption{(a) Final population of the output qubit as a function of the weight $w_1 = -J$ for two different bias values $b/2\pi$ of $0.8\,\mathrm{MHz}$ and $4.0\,\mathrm{MHz}$. As expected, when the input qubit is in the $|0\rangle$ state, the final output state does not depend on the weight, while for control state $|1\rangle$ it follows the nonlinear activation function. The solid and dashed lines are derived from numerical simulations with no fit parameters. (b) Negativity of the final two-qubits state after preparing the input qubit in an equal superposition state and the output qubit in its ground state and applying the perceptron gate with a weight $w_1/2\pi = -5.2\,\mathrm{MHz}$. Negativity of a two-qubit state can range between $0$ for a separable state to $1/2$  for a maximally entangled state. Lines show the simulated negativity for perfect unitary dynamics (solid) and when including relaxation of the qubits with decay times of $T_1=20\,\mathrm{\mu s}$ (dashed).}\label{fig:maxdiff}
\end{figure}

As the next step towards trainability of the network, we control the weight $w_1$ at a fixed bias (see Fig.~\ref{fig:maxdiff}(a)). For the input qubit in the ground state, the final state of the output does not depend on $w_1$, since $x_1 = 0$. For the control in the excited state, the dependence on $w_1$ is simply the activation function itself, shifted by choosing different values for the bias $b$.

We further characterize the perceptron gate with quantum process tomography \cite{Mohseni2008} and verify several important aspect of the implemented process with the extracted process matrices: It is expected to act as a controlled gate -- that is, if the input qubit is prepared in one of the basis states $|0\rangle$, $|1\rangle$, the gate should leave it in this state, independently of the output qubit's initial state. We calculate the average fidelity with which the process $\mathcal{M}$ satisfies this condition, i.e.
\begin{equation}\label{eq:mfid}
  \overline{\mathcal{F}} =
  \frac{1}{2}\sum_{i\in\{0,1\}}
  \Big\langle
  \mathrm{Tr}
  \big(|i\rangle\langle i|\otimes \mathbbm{1}
  \big)
  \mathcal{M}\big[|i\rangle\langle i|\otimes 
  |\varphi\rangle\langle\varphi|\big]
  \Big\rangle_{\varphi},
\end{equation}
where $\langle\cdot\rangle_{\varphi}$ denotes averaging over the initial state $|\varphi\rangle$ of the input qubit. This average is calculated from the process matrix of $\mathcal{M}$ by directly using Eq.~(\ref{eq:mfid}) and the identity $\big\langle|\varphi\rangle\langle\varphi|\big\rangle_{\varphi} = \mathbbm{1}/2$. The obtained fidelity values are independent of the detuning $\Delta$ and lie between 0.95 and 0.97, indicating that the implemented operation is to a good approximation a controlled gate. We also evaluate the purity of the final state averaged over the initially prepared state and obtain a value around 0.78 (for a $1.7\,\mu\mathrm{s}$ long pulse), independent of the qubit-drive detuning. This confirms that the gate is mostly a unitary process, limited slightly by qubit decoherence. The average purity value is consistent with typical observed qubit dephasing times which vary in our experiment between $10~\mu\rm{s}$ and $20~\mu\rm{s}$.

Finally, to show the entangling property of the perceptron gate, we prepare the input qubit in an equal superposition state and the output qubit in its ground state. We then apply the perceptron gate, extract the density matrix of the final two-qubit state and calculate its negativity \cite{Vidal2002}, which reaches nearly the maximum value of $1/2$ for bias values at which the activation functions for the two computational states of the input qubit are well separated (Fig.~\ref{fig:maxdiff}(b)). 
In the limit of large weight, the ideal operation at the mid-point between the activation function slopes would become equivalent (up to local phases) to a controlled NOT gate, preparing a Bell state and reaching maximum negativity.

\begin{figure}
    \centering
    \includegraphics[width=8.5cm]{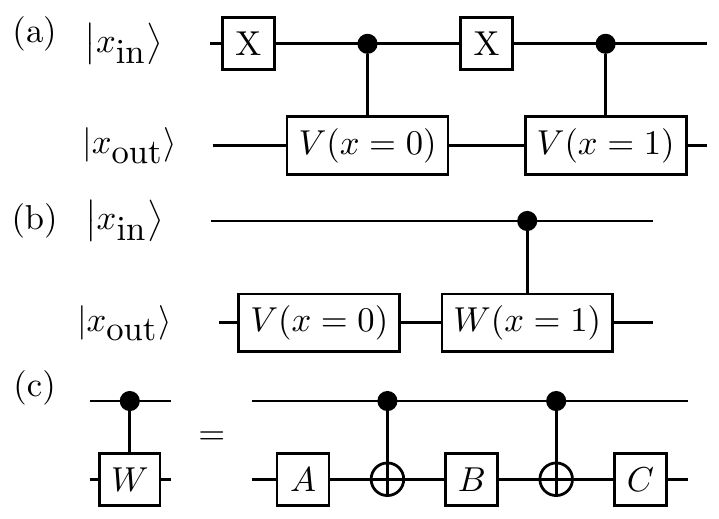}
    \caption{(a) A circuit that implements the equivalent of the perceptron gate with a single input and a single output qubit.
    A controlled-$V$ operation needs to be applied for each possible string of the inputs, in this case $x_\text{in} \in \{0,1\}$.
    (b) The circuit can be simplified by applying one of the multi-qubit controlled operations as a single-qubit gate and adjusting the following operations. Here, $W(x=1)=V(x=1) \cdot V(x=0)^\dagger$.
    (c) The controlled-$W$ gate is further decomposed into at most two CNOT gates and three single qubit gates $A$, $B$ and $C$.
    \label{fig:AR_eq_2q} }
\end{figure}

\section{Equivalent circuit and \\ scaling complexity}\label{sec:equiv-circuit}
For a general, multi-qubit input register the perceptron's action in Eq.~(\ref{eq:perceptrongate}) can be considered as a generalized multi-qubit conditional gate:
A separate operation $V(x)$ is applied for each of the basis states $\ket{x}\equiv\ket{x_1x_2\ldots x_N}$ of the $N$ input qubits. This is in contrast to a single multi-qubit controlled operation, that applies the operation $V$ to a output qubit only if all input qubits are in the state $\ket{1}$.
Thus, to implement the perceptron gate requires a multi-qubit controlled gate for each input basis state, in total $2^{N}$ operations. 
This can be reduced to $2^{N}-1$ by implementing the $V(x)$ of one of the input strings as a single qubit gate and adjusting all multi-qubit controlled operations accordingly.
For small $N\lesssim 10$, a multi-qubit controlled gate can be decomposed into $2^{N+1}-2$ two-qubit gates \cite{Barenco1995}, which gives us a total gate count for the equivalent circuit of $N_g = (2^N -1)(2^{N+1} - 2)$, approaching  $\sim 2^{2N+1}$ for large number of input qubits.

More specifically, the single-input perceptron implemented in this paper can be decomposed into single qubit gates and two CNOT gates (Fig.~\ref{fig:AR_eq_2q}).The fidelity and total duration of a sequence is typically limited by the number of two-qubit gates in transmon-based architectures. Hence, when ignoring single qubit gates, the sequence equivalent to the perceptron gate would be implemented in a gate time of $\sim120~\rm{ns}$ and with an estimated fidelity of $\sim99.4\%$. Here, we have assumed state of the art CNOT/CZ gates with fidelities of up to $99.7\%$ \cite{kjaergaard2020_gate, Kjaergaard2020} and gate times of $60\rm{ns}$.
The advantage of the adiabatic protocol, with approximately constant time, over the equivalent unitary circuit becomes apparent as the number of input qubits is increased, as discussed in Appendix~\ref{app:EquivalentUnitary}. For example, when decomposing the perceptron gate with two input qubits the best scenario estimate suggests a fidelity of $\sim94.7\%$ with a gate time of $\sim1.1~\mu\textrm{s}$.

\section{Discussion and Outlook}
By realizing the adiabatic perceptron dynamics that activates the output qubit depending on the state of connected input qubits directly in hardware, we have demonstrated the basic building block of a quantum feed-forward neural network. In this co-design approach, we show that by changing the length of the perceptron adiabatic qubit pulse the shape of the activation function can be modified, whereas the ZZ shift mediated by a tunable coupler and the adiabatic drive final frequency can be tuned to modify the weights and bias, which allows for training of the neural network.

As an extension of the demonstrated single-digit input scenario, an efficient implementation of a multi-input quantum perceptron can be realized by coupling multiple input qubits, each with its own tunable coupler, to a single output qubit. Its cumulative frequency shift would be of the form shown in Eq.~(\ref{eq:zzintham}) \cite{Nigg2013}. It is noteworthy that the adiabatic gate presented here leads to complex multi-qubit operations with two-body interactions only and with a gate time that is independent of the number of inputs. In contrast, the time to run the equivalent circuit scales exponentially in the number of qubits, with significant advantage of the co-designed perceptron expected already for the three-qubit input case. 
While the quantum perceptron we co-design exponentially reduces the number of required conventional gates, all-to-all connectivity between the qubits of adjacent layers is desired to fully benefit from this advantage. Limitations imposed by the number of couplers that can be physically attached to a single  qubit or compromises in the connectivity between layers may reduce this advantage. Therefore, multi-qubit couplers \cite{Mezzacapo2014,Chancellor2016,Paik2016,Sameti2017} may become increasingly useful, as will be architectures that combine the presented adiabatic implementation of a perceptron with standard digital gate sequences.

To ensure proper trainability of the network, the range of achievable weigths and biases should allow to probe the activation function at both limits $f(x)\approx1$ and
$f(x)\approx0$. For our implementation, the possible range of ZZ shifts should be comparable to the charateristic width of the adiabatic curve, that is in turn limited by the ability to follow the eigenstates adiabatically given by the speed of the gate.
Therefore, to reduce the perceptron gate time one could envisage either non-adiabatic versions of the perceptron \cite{Ban2021}, or an extended range of achievable ZZ coupling by AC \cite{Noguchi2020, Ganzhorn2019} or DC \cite{Xu2021} pulses on the tunable coupler.
 
While we have demonstrated the entangling power of the perceptron gate, evaluating quantum advantage in larger networks is subject to forthcoming investigations.
For instance, it will be important to understand how decoherence will affect the result and what role the entanglement between the input and output layer plays, in particular if the input layer is in a superposition state. Partially tracing out input and hidden layers and recycling qubits might allow for larger networks and protect the results from decoherence of earlier layers.
Moreover, an investigation of the effects of higher connectivity on the expressivity of the nextwork will be required.

Finally, neural networks based on adiabatic perceptron gates may also have applications not directly related to machine learning. As an example, it is plausible that their quantum counterparts could serve well to parametrize unitaries in variational quantum algorithms \cite{ Farhi2014,Peruzzo2014,Kandala2017}, similar to the efficient approximation capability found in classical neural networks.

\section{Acknowledgments}

We thank Stephan Paredes, Andreas Fuhrer, Matthias Mergenthaler, Peter M\"{u}ller and Clemens M\"{u}ller for insightful discussions and the quantum team at IBM T. J. Watson Research Center, Yorktown Heights for the provision of qubit devices. We thank R. Heller and H. Steinauer for technical support.
Fabrication of samples was financially supported by the ARO under contract W911NF-14-1-0124, F.R. and M. W. acknowledge funding by the European Commission Marie Curie ETN project QuSCo (Grant Nr. 765267), and M. P., S. W. and G. S. by the European FET-OPEN project Quromorphic (Grant Nr. 828826).

\bibliography{references}

\appendix

\begin{figure*}[t!]
    \centering
     \includegraphics[width=14cm]{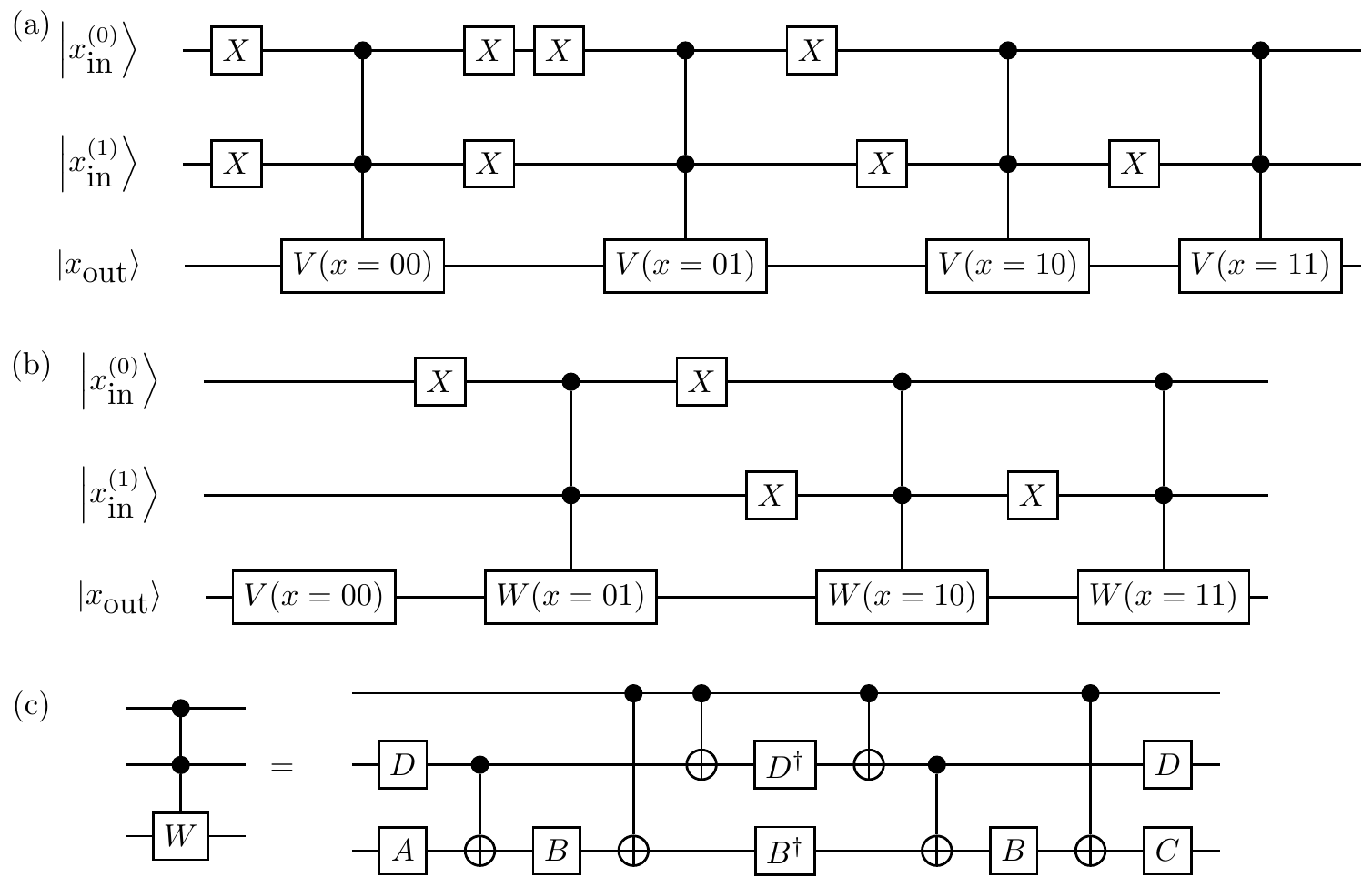}
    \caption{
    (a) Decomposition of the perceptron gate with two inputs and a single output qubit into controlled-controlled-$V$ operations and single-qubit operations $X$.
    (b) Improved circuit decomposition: one of the controlled-controlled operations, $V(00)$, is applied as a single-qubit gate with a simple adjustment $V(x)\rightarrow W(x)=V(x)V(00)^\dagger$ for the other controlled-controlled operations.
    (c) Decomposition of a controlled-controlled-$W$ gate in terms of 6 CNOT gates and 8 single-qubit gates. \label{fig:AR_eq_3q} }
\end{figure*}

\section{Equivalent unitary}\label{app:EquivalentUnitary}
The equivalent unitary matrix for the adiabatic perceptron can be expressed in terms of standard gates such a two-qubit controlled NOT and single qubit gates. However, the depth of the equivalent circuit grows exponentially with the number of input qubits.

The action of the perceptron in Eq.~(\ref{eq:perceptrongate}) can be thought of as a generalized multi-conditional gate. Whereas a `conventional' N-controlled-$V$ operation applies the gate $V$ to a output qubit only if the control qubit (or qubits, for a multi-qubit controlled gate) is in the $\ket{1}$ state, in the perceptron gate for every computational basis state $\ket{x}\equiv\ket{x_1 x_2\ldots x_N}$ with $x_i \in \{0,1\}$ of the control qubits a different gate $V(x)$ is applied to the output qubit. The matrix representation for such a controlled-gate has a block-diagonal structure
\begin{equation*}
U = \begin{pmatrix}
V(x=0\dots0) & & & \\
& V(x=0\dots1) & & \\
& & \ddots & \\
& & & V(x=1\dots1)
\end{pmatrix}
\end{equation*}
with each block $V(x)$ corresponding to a particular input string $x$.
Thus, the perceptron gate can be expressed in terms of multi-qubit controlled unitaries as illustrated for a single input qubit in Fig.~\ref{fig:AR_eq_2q}(a) and for two input qubits in Fig.~\ref{fig:AR_eq_3q}(a). However, the number of multi-qubit controlled unitaries can be decreased by one by applying one the $V(x)$ gates as a single qubit gates, e.g. the unitary $V(x=0\dots0)$ conditioned on all inputs being in the ground state, and adjusting all other multi-qubit controlled gates so as to account for the extra operation, in this example $V(x)\rightarrow W(x)=V(x)V(0\dots0)^\dagger$. See Fig.~\ref{fig:AR_eq_2q}(b) and Fig.~\ref{fig:AR_eq_3q}(b) for one- and two-input qubits examples.

When acting on an output qubit initialized in the $\ket{0}$ state, the perceptron gate corresponds to the map
\begin{equation}
\ket{x}\otimes\ket{0} \rightarrow \ket{x}\otimes\left( \sqrt{1 - |c(x)|^2}\ket{0} + c(x)\ket{1}\right).
\end{equation}
By imposing unitarity, we can infer the action of the perceptron gate on a output qubit in the $\ket{1}$ state, and the controlled-$V(x)$ gate with
\begin{equation}
V(x) = \begin{pmatrix}
\sqrt{1 - |c(x)|^2} & c^*(x)\\
-c(x) & \sqrt{1 - |c(x)|^2}.
\end{pmatrix}.
\end{equation}

For real $c(x)$, we can write $c(x) = \sin\theta_x/2$ and $\sqrt{1 - c^2(x)} = \cos\theta_x/2$ and express $V(x)$ as
\begin{equation}
V(x) = \begin{pmatrix}
\cos\theta_x/2 & \sin\theta_x/2\\ -\sin\theta_x/2 & \cos\theta_x/2
\end{pmatrix} = R_y(\theta_x).
\end{equation}

That is to say, the perceptron reduces to a sequence of rotations around the $X, Y$ and $Z$ axes.
The number of multi-qubit controlled rotations in this decomposition is equal to the number of possible input bitstrings, which scales exponentially with the number of input qubits.
As discussed in the main text, this may lead to a potential advantage of this protocol over more standard paramaterized gates.

Moreover, near-term quantum computers cannot generally implement multi-qubit controlled gates at once, but rather must decompose them into a series of one- and two-qubit gates as shown in ~Figs.~\ref{fig:AR_eq_2q}(c)~and~\ref{fig:AR_eq_3q}(c) by using CNOT gates as the primitive two-qubit operation.

In the main text, we use gate times and fidelities quoted for CZ instead of CNOT gates, being $F=99.7\%$ and $\tau=60~\rm{ns}$ as these are currently better for state-of-the-art transmon qubit architectures and since a CZ can easily be transformed to a CNOT via single qubit gates.

With these decompositions, and assuming that the two-qubit gates are the dominant source of error, we obtain estimates for the equivalent fidelities and gate times quoted in the main text:
for the two-qubit (single-input, single-output) gate, 2 CNOTs are needed, leading to a fidelity of $\sim98\%$ and a gate time of $\sim240~\rm{ns}$;
for the three-qubit (two inputs, one output) gate 18 CNOTs are needed, leading to a fidelity of $\sim91\%$ and a gate time of $\sim1.1~\mu\textrm{s}$. Gate time and fidelity estimates  for more inputs are summarised in Table~\ref{table:scaling}.

\begin{table}[t]
\begin{tabular}{ c c c c c } 
\hline
 $N$ & 1 & 2 & 3 & 4 \\ \hline
 $N_g$ & 2 & 18 & 98 & 450 \\ \hline
 $t$ ($\mu$s) & 0.12 & 1.1 & 5.9 & 27 \\ \hline
 $\mathcal{F}$ & 0.994 & 0.95 & 0.75 & 0.26 \\
 \hline
\end{tabular}
\caption{\label{table:scaling} Number of CNOT gates $N_g$ required for a circuit equivalent to the adiabatic ramp protocol with $N$ input qubits, along with estimates for the total circuit time $t$ and the circuit fidelity $\mathcal{F}$.}
\end{table}

\section{Hamiltonian and Perturbation theory}\label{app:pertThJ}

The Hamiltonian describing the system is 
\begin{equation}
    \begin{aligned}
    H / \hbar = \sum_{i=1,2,c} &\omega_i a_i^\dagger a_i + \frac{\alpha_i}{2} a_i^\dagger a_i^\dagger a_i a_i\\
    &+ \sum_{i=1,2} g_{ic} (a_i - a_i^\dagger)(a_c - a_c^\dagger)
    \end{aligned}
    \label{eq:ham}
\end{equation}

where $a_i^\dagger$ ($a_i$) are creation (annihilation) operators, $\omega_i$ are the frequencies and $\alpha_i$ are the anharmonicities of qubits ($i=1,2$) and coupler ($i=c$), and $g_{ic}$ are their respective couplings. 
\begin{figure}[h]
	\includegraphics[width=0.38\textwidth]{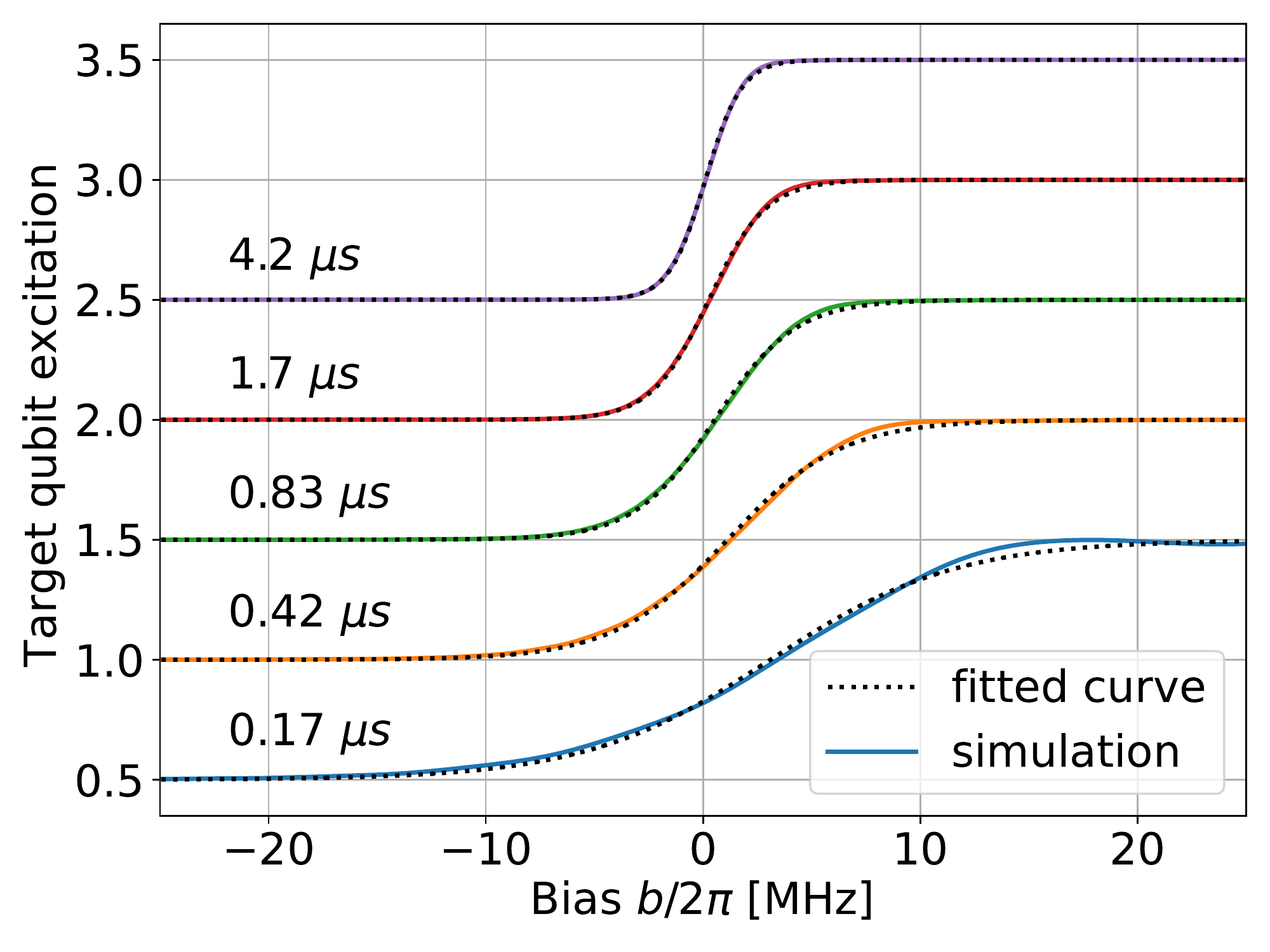}
	\centering
	\caption{\label{fig:analytical_chirped} Comparison for different gate times of the simulated populations, solid line, and fits to the analytical (Eq. \ref{eq:analytical_chirped}) from \cite{Hioe1984}. The analytical form well describes the simulated results, with small deviations for short gate times.
	}
\end{figure}
The anharmonicities $\alpha_i$ are defined as the difference $\omega_{12}-\omega_{01}$ between the $0\leftrightarrow 1$ and $1\leftrightarrow 2$ transition frequencies; values for the measured parameters can be found in \cite{Ganzhorn2019}.

Using fourth-order perturbation theory, we can approximate the ZZ coupling strength $J$ as

\begin{widetext}
\begin{equation}\label{eq:pertThJ}
  J =
  2 g_{1c}^2 g_{2c}^2
  \frac{
  \alpha_1\alpha_2(\Delta_1+\Delta_2)^2 + \alpha_2\alpha_c\Delta_1^2 + \alpha_1\Delta_1^2(\Delta_1+\Delta_2) + \alpha_2\Delta_2^2(\Delta_1+\Delta_2) + \alpha_c(\Delta_1+\Delta_2)(\Delta_1-\Delta_2)^2
  }{
  \Delta_1^2 \Delta_2^2 (\Delta_1+\Delta_2+\alpha_c)(\Delta_2-\Delta_1-\alpha_2)(\Delta_1-\Delta_2-\alpha_1),
  }
\end{equation}
\end{widetext}
where $\Delta_{1,2} = \omega_{c}-\omega_{1,2}$ are the detunings of the coupler from the two qubits.

\section{Analytic solution for chirped pulses}\label{app:chirped}
The driven two level system, although ubiquitous, has proven to notoriously difficult to solve analytically. The most well known analytical solution is the Landau-Zener linear ramp \cite{landau1932,zener1932}. Soon after another exact solution was found for a hyperbolic secant pulse \cite{rosen1932} and was later extensively used \cite{mccall1969}, analysed and extended \cite{Hioe1984}.
The hyperbolic secant pulse is chirped to have a detuning $\omega_{\mathrm{p}}(t)$ and modulated in amplitude $\Omega(t)$ according to:

\begin{equation}
\begin{aligned}
\omega_{\mathrm{p}}(t) &= \omega_{\mathrm{i}} + (\omega_{\mathrm{f}}-\omega_\mathrm{i})\tanh^2(\pi t / T),
\\
\Omega(t) &= \Omega_0 \sech(\pi t/T).
\end{aligned}
\end{equation}
Assuming that the pulse begins at time $t=-\infty$ and ends at time $t=+\infty$ one can find an analytical expression for the population transfer at the end of the pulse. This is given as
\begin{widetext}
\begin{equation} \label{eq:analytical_chirped}
\begin{aligned}
    P_1(+\infty) = \sech\left[(\omega_i+\Delta_f)T/2\right]  \sech\left[(\omega_i-\Delta_f)T/2\right]
    \left[ \sin^2(\sqrt{\Omega_0^2 + \Delta_f^2}T/2) +  \sinh^2(\Delta_f T/2)
    \right]
\end{aligned}
\end{equation}
\end{widetext}

The chirped pulses in our experiment can be obtained from the ones described above by applying the time transformation
\begin{align*}
t \rightarrow \mathrm{artanh}(-\cos(\pi*t'/T))*T
\end{align*}
and indeed define the same trajectory in the detuning versus amplitude plane.
When we compare the analytical solution described above with the simulation results we find agreement in the curve behaviour (Fig.~\ref{fig:analytical_chirped}). Differences in steepness of the curves are due to the time transformation between the two cases. Nonetheless the underlying sigmoid behaviour remains, and the curves can be precisely matched by allowing for the fitting parameter $T$ and a further detuning of the whole curve $\delta$, given by $\omega_i \rightarrow \omega_i + \delta$.

\end{document}